\documentclass[useAMS,usenatbib]{mn2e}

\usepackage[utf8x]{inputenc}
\usepackage{graphicx}
\DeclareMathAlphabet{\mathpzc}{OT1}{pzc}{m}{it} 

\title[A comparison of CMB TT APS Estimators at Large Scales]{A comparison of CMB Angular Power Spectrum Estimators at Large Scales: the TT case}

\author[D. Molinari et al.]
{\parbox{\textwidth}{D. Molinari$^{1,2}$\thanks{E-mail:molinari@iasfbo.inaf.it}, A. Gruppuso$^{2,3}$, G. Polenta$^{4,5}$, C. Burigana$^{2,6}$, A. De Rosa$^{2}$, \\ P. Natoli$^{6,7,4,2}$, F. Finelli$^{2,3}$, and F. Paci$^{8}$}\vspace{0.4cm}\\
\parbox{\textwidth}{$^{1}$Dipartimento di Fisica e Astronomia, Universit\`a degli Studi di Bologna, viale Berti Pichat 6/2, I-40127 Bologna, Italy\\
$^{2}$INAF-IASF Bologna, Via Piero Gobetti 101, I-40129, Bologna, Italy\\
$^{3}$INFN, Sezione di Bologna, Via Irnerio 46, I-40126 Bologna, Italy \\
$^{4}$Agenzia Spaziale Italiana Science Data Center, c/o ESRIN, via Galileo Galilei, Frascati, Italy \\
$^{5}$INAF - Osservatorio Astronomico di Roma, via di Frascati 33, Monte Porzio Catone, Italy \\
$^{6}$Dipartimento di Fisica e Scienze della Terra, Universit\`a degli Studi di Ferrara, via Giuseppe Saragat 1, I-44100 Ferrara, Italy\\
$^{7}$INFN, Sezione di Ferrara, via Giuseppe Saragat 1, I-44100 Ferrara, Italy\\
$^{8}$SISSA - Scuola Internazionale Superiore di Studi Avanzati, via Bonomea 265, I-34136 Trieste, Italy}}

\begin{document}

\date{5 March 2014}

\pagerange{\pageref{firstpage}--\pageref{lastpage}} \pubyear{2014}

\maketitle

\label{firstpage}

\begin{abstract}
In the context of cosmic microwave background (CMB) data analysis, we compare the efficiency at large scale  of two angular power spectrum algorithms,  implementing, respectively, the quadratic maximum likelihood (QML) estimator and the pseudo spectrum (pseudo-$C_\ell$) estimator. By exploiting 1000 realistic Monte Carlo (MC) simulations, we find that the QML approach is markedly superior in the range $2 \leq \ell \leq 100$. At the largest angular scales, e.g. $\ell \leq 10$, the variance of the QML is almost $1/3$ ($1/2$) that of the pseudo-$C_\ell$, when we consider the WMAP kq85 (kq85 enlarged by 8 degrees) mask, making the pseudo spectrum estimator a very poor option. Even at multipoles $20 \leq \ell \leq 60$, where pseudo-$C_\ell$ methods are traditionally used to feed the CMB likelihood algorithms, we find an efficiency loss of about $20\%$, when we considered the WMAP kq85 mask, and of about $15\%$ for the kq85 mask enlarged by 8 degrees. This should be taken into account when claiming accurate results based on pseudo-$C_\ell$ methods. Some examples concerning typical large scale estimators are provided.
\end{abstract}


\section{Introduction}

The pattern of the cosmic microwave background (CMB) anisotropy field can be used to probe cosmology to high precision, as shown by the  Wilkinson Microwave Anisotropy Probe (WMAP) 9 years results 
\citep{Hinshaw:2012fq} and by the very recent {\it Planck} cosmological results (see \cite{Ade:2013xsa} and references therein).  
CMB data have given a significant contribution in setting up the $\Lambda$ cold dark matter ($\Lambda$CDM) cosmological concordance model. 
The latter establishes a set of basic quantities for which CMB observations and other cosmological and astrophysical data-sets 
agree\footnote{See, however, \cite{Ade:2013lmv} for a possible tension concerning the  $\Omega_m$ estimate  from {\it Planck} CMB and galaxy clusters data.}: spatial curvature close to zero;  $\sim 68.5 \%$ of the cosmic density in the form of Dark Energy; $\sim 26.5\%$ in cold dark matter; $\sim 5\%$ in baryonic matter; and non perfectly scale invariant adiabatic, primordial perturbations compatible with Gaussianity \citep{Ade:2013lta,Ade:2013ydc}.

In particular, the largest scales of the temperature anisotropies map are of great interest because they directly probe the Early Universe (or Inflationary Phase of the Universe \cite{Starobinsky:1980te,Guth:1980zm,Linde:1981mu,Albrecht:1982wi}).
They correspond to angular scales larger than the horizon size at decoupling as observed today, i.e. $\theta  >  2^{\circ}$ or, equivalently $\ell < \ell_{dec} \sim 90$ (see for example \cite{Page2003})
with $\ell $ being the multipole order of the spherical harmonics expansion
\begin{equation}
{\delta T}(\hat n) = \sum_{\ell m} a_{\ell m} \, Y_{\ell m} (\hat n) 
\label{SHE} \, ,
\end{equation}
where $\delta T = T(\hat n) - T_0$ is the temperature anisotropy observed in the direction $\hat n$ relative to the CMB average temperature $T_0 \simeq 2.725 K$ (\cite{Mather1999}; see also \cite{Fixsen1996} for a constraint of the CMB black body shape), and
with $a_{\ell m}$  being the coefficients of the Spherical Harmonics $Y_{\ell m} (\hat n) $.
 
The main contribution to the CMB anisotropies at these large scales is provided by the so called Sachs-Wolfe effect and by a subdominant integrated Sachs-Wolfe effect \citep{Sachs:1967er}
which is different from zero because of the recent (from a cosmological point of view) transition to an accelerated phase of the Universe \citep{Kofman:1985fp} likely associated to a dark energy (or Cosmological constant) component. 
In principle a stochastic background of primordial gravitational waves can also give a contribution to the temperature CMB anisotropies at these largest scales, depending on the tensor-to-scalar ratio, $r$, constrained by the current data \citep{Hinshaw:2012fq,Ade:2013lta}. A firm detection of the primordial gravitational waves requires CMB B modes polarization measurement at large scale \citep{Knox:1994qj}.

From the observational point of view the CMB anisotropies temperature map, as observed by WMAP 9 year, is cosmic variance dominated, i.e., cosmic variance exceeds the instrument noise, up to $\ell = 946$,
see \citep{Bennett:2012fp}. For {\it Planck} data this crossing happens at $\ell \sim 1500$ \citep{Planck:2013kta}.
Therefore, at the largest scales ($\ell < \ell_{dec}$), the effect of instrumental noise is almost negligible. 
Keeping this in mind, it is even more important to employ the most accurate data analysis tools.

In the current paper we focus on the angular power spectrum (APS), which is the main observable for diagnosis of the CMB map.
The method that is capable to provide APS with no bias and with the minimum variance, as provided by the Fisher-Cramer-Rao inequality is the Quadratic Maximum Likelihood (QML) method \citep{tegmark_tt,tegmark_pol}. Such an optimal method has the drawback of being computationally expensive and then limited by the number of pixels. It is currently implemented and applied at low resolution (see e.g. \cite{Gruppuso2009}).
Several other strategies for measuring $C_{\ell}$ at low resolution have been developed and applied to CMB data with excellent results. These methods include different sampling techniques such as Gibbs \citep{jewell, wandelt, eriksen}, adaptive importance \citep{teasing} and Hamiltonian \citep{taylor}.
At high multipoles ($\ell > 30$, \cite{efstathiou}) the so called pseudo-$C_{\ell}$ algorithms are usually preferred to others techniques.
These methods, in fact, implement the estimation of power spectral densities from periodograms \citep{hauser}.
Basically, they estimate the $C_l$ through the inverse Harmonical transform of a masked map that is then deconvolved with geometrical kernels and corrected with a noise bias term.
These techniques, such as Master \citep{master}, Cross-Spectra \citep{saha,polenta,grain}, give unbiased estimates of the CMB power spectra and moreover, it has been shown they work successfully when applied to real data at high multipoles \citep{ACBAR,Jones:2006,MAXIPOL,dunkley_wmap5}.
These estimators are pretty quick and light from a computational point of view.
However, it is well known that at low multipoles they are not optimal since they provide power spectra estimates with error bars larger than the minimum variance.
We note that in \cite{efstathiou2} an hybrid approach has already been proposed, i.e. the QML at low and the pseudo-$C_{\ell}$ at high multipoles, where a recipe for an hybrid covariance is consistently given. However, in that paper the QML is considered only up to ${\ell}=40$ since it is applied to maps with a full width at half maximum (FWHM) $=3 ^{\circ}$.

The aim of the present work is to compare quantitatively the QML and pseudo-$C_{\ell}$ methods under {\it realistic} assumptions focusing on the largest scales (i.e. low multipoles) where CMB anomalies are mainly located \citep{anomalies,Ade:2013sta}.
The idea is to provide the scientific community with a reference analysis comparing the heavy QML method and the light and quick pseudo-$C_{\ell}$ approach in the range of multipoles $\ell < \ell_{dec} \sim 90$.
Here we quantitatively address this problem through realistic Monte Carlo (MC) simulations.
A similar analysis has been carried out by WMAP team using the so-called $C^{-1}$ method employed in \cite{Bennett:2012fp} that shows an improvement with respect to the pseudo-$C_{\ell}$ method mostly located at very low multipoles and in the intermediate S/N regime. Moreover, we evaluate the benefit of considering an optimal APS estimation taking into account some examples of estimators of anomalies that are commonly used at large scales to test the consistency of the observations with the $\Lambda CDM$ model.

The paper is organized as follows.
In Section \ref{descriptionscodes} we describe the considered implementations of QML and pseudo-$C_{\ell}$ codes, called respectively {\it BolPol}
and {\it cROMAster}.
In Section \ref{computationalrequirements}  the computational requirements for both the methods are given.
Section \ref{comparison} is devoted to the detailed description of the comparison and corresponding results. 
Examples on how the previous analysis propagates to the APS based estimators, commonly used in literature for the analysis of the large scales anomalies, are provided in Section \ref{WMAP}.
Conclusions are drawn in Section \ref{conclusions}.

\section{Description of the codes}
\label{descriptionscodes}

In this section we give a description of the two considered APS estimators. Specifically we consider {\it cROMAster}, as an implementation of pseudo-$C_{\ell}$ method and {\it BolPol} as implementation of the QML estimator.

\subsection{BolPol}
\label{bolpol}

In order to evaluate the APS of full sky and masked sky maps {\it BolPol} code adopts the QML estimator, introduced in \cite{tegmark_tt}. 
In this section we describe the essence of this method.
Given a CMB temperature map $\textbf{x}$, the QML provides estimates $\tilde{C}_{\ell}$ of the TT APS as

\begin{equation}
\hat{C}_{\ell}= \sum_{\ell'} (F^{-1})_{\ell\ell'} \left[ \textbf{x}^t\textbf{E}^{\ell'}\textbf{x}-tr(\textbf{NE}^{\ell'})\right] ,
\end{equation}
where $F_{\ell\ell'}$ is the Fisher matrix, defined as

\begin{equation}\label{FF}
F_{\ell\ell'} = \frac{1}{2} tr \left[ \textbf{C}^{-1} \frac{\partial{\textbf{C}}}{\partial{C_{\ell}}} \textbf{C}^{-1} \frac{\partial{\textbf{C}}}{\partial{C_{\ell'}}} \right] ,
\end{equation}
and the $\textbf{E}^{\ell}$ matrix is given by

\begin{equation}\label{El}
\textbf{E}^{\ell} = \frac{1}{2} \textbf{C}^{-1} \frac{\partial{\textbf{C}}}{\partial{C_{\ell}}} \textbf{C}^{-1} ,
\end{equation}
with $\textbf{C}=\textbf{S}(C_{\ell})+\textbf{N}$ being the global covariance matrix (signal plus noise contribution).

Although an initial assumption for a fiducial power spectrum $C_{\ell}$ is needed in order to build the signal covariance matrix $\textbf{S}(C_{\ell})$, it has been proven that the QML method provides unbiased estimates of the power spectrum contained in the map regardless of the initial guess \citep{tegmark_tt},

\begin{equation}
\langle \hat{C}_{\ell} \rangle = \overline{C}_{\ell} \, ,
\label{eq:unbiased}
\end{equation}
where the average is taken over the ensemble of realizations (or, in a practical test, over MC realizations extracted from $\overline{C}_{\ell}$). 
On the other hand, the covariance matrix associated to the estimates is given by the inverse of the Fisher matrix

\begin{equation}
\langle \Delta\hat{C}_{\ell} \Delta\hat{C}_{\ell'} \rangle = (F^{-1})_{\ell\ell'} \, ,
\label{eq:error bars}
\end{equation}
only when the fiducial spectrum $C_{\ell} = \overline{C}_{\ell}$.
In this case the inverse of the Fisher matrix provides the minimum variance.
According to the Cramer-Rao inequality, which sets a limit to the accuracy of an estimator, equation (\ref{eq:error bars}) tells us that the QML has the smallest error bars. 
The QML is then an {\it optimal} estimator because equations (\ref{eq:unbiased}) and (\ref{eq:error bars}) are both satisfied.

For a generalization to polarization spectra see \cite{tegmark_pol}. 

\subsection{cROMAster}
\label{cromaster}

{\it cROMAster} is an implementation of the pseudo-$C_{\ell}$ method proposed by \citet{master}, extended to allow for both auto- and cross-power spectrum estimation
(see \citet{polenta} for a comparison).
A given sky map $\textbf{x}$ can be decomposed in CMB signal and noise as $\textbf{x}=\textbf{s}+\textbf{n}$, and we define the masked sky pseudo-spectrum as:
\begin{equation}
	\tilde{C}_{\ell} = \frac{1}{2\ell +1}\sum_{m=-\ell}^{\ell} |\tilde{a}_{\ell m}|^2 = \tilde{C}_{\ell}^{S} + \tilde{N}_{\ell} \, ,
\end{equation}
where $\tilde{C}_{\ell}^{S}$ is the pseudo-spectrum of the sky signal, $\tilde{N}_{\ell}$ is the pseudo-spectrum of the noise present in the map,
and the pseudo-$a_{\ell m}$ coefficients are computed as:
\begin{equation}
	\tilde{a}_{\ell m} = \int d\Omega x(\theta,\phi) w(\theta,\phi) Y_{\ell m}^{*}(\theta,\phi) \, ,
\end{equation}
where $\textbf{w}$ is the applied mask.

In order to recover the full sky power spectrum, we employ the following estimator:
\begin{equation}\label{eq:pseudocl}
	\hat{C}_{\ell} = K_{\ell \ell^{'}}^{-1} \left( \tilde{C}_{\ell} - \tilde{N}_{\ell} \right)
\end{equation}
where the mode-mode coupling kernel $K_{\ell \ell^{'}}^{-1}$ is a geometrical correction that accounts for the loss of orthonormality of the spherical harmonic functions in the cut sky
(see \citet{master} for more details). In fact, for an isotropic sky signal that is a realization of a theoretical power spectrum $C_{\ell}$, one can write:
\begin{equation}
	\langle \hat{C}_{\ell} \rangle = C_{\ell} \, .
\end{equation}
Hence, equation (\ref{eq:pseudocl}) defines an unbiased estimator provided that the noise term $\tilde{N}_{\ell}$ is properly removed, which is usually done through MC simulations. However, for current generation 
high-sensitivity experiments, such as {\it WMAP} and {\it Planck}, the noise bias has to be known to better than $0.1\%$ accuracy, 
and therefore the cross-spectrum approach is preferred (see, e.g. \citet{Bennett:2012fp, Planck:2013kta}). 
In the latter case, the pseudo-$C_{\ell}$ are computed by combining the spherical harmonic coefficients from two or more maps with uncorrelated noise
so that the ensemble average of $ \tilde{N}_{\ell}$ is null, making the cross-spectrum estimator naturally unbiased.

For practical applications, i.e. involving colored noise properties and arbitrary masks incorporating possible non-uniform weighting of the data, there is no readily available exact analytical expression for the covariance matrix of this estimator. 
The error bars for pseudo-$C_{\ell}$ methods can be roughly approximated as:
\begin{equation}\label{eq:rot}
	\Delta \hat{C}_{\ell} = \sqrt{\frac{2}{(2\ell + 1)f_{sky}^{eff}}}\left( C_{\ell} + N_{\ell} \right)
\end{equation}
where $f_{sky}^{eff}$ is the effective fraction of the sky used for the analysis which accounts for the weighting scheme of the pixels. For a given sky coverage, a uniform weighting scheme produces the largest 
$f_{sky}^{eff}=f_{sky}$ and is therefore the optimal choice in the signal dominated regime, which corresponds to large angular scales. On the other hand, an inverse noise weighting scheme reduces $N_{\ell}$ and is the
best choice in the noise dominated regime, i.e. at small angular scales. A trade off between the two weighting scheme should be applied in the intermediate regime, and an optimal pseudo-$C_{\ell}$ method should employ the best weighting scheme at a given multipole according to the S/N ratio at that multipole.

In order to derive an accurate estimate of the covariance matrix, the original approach proposed in \citet{master} is to rely on signal and noise MC simulations:
\begin{equation}\label{eq:mccov}
	Cov \left\{ \hat{C}_{\ell}, \hat{C}_{\ell^{'}} \right\} = \langle ( \hat{C}_{\ell} - \langle \hat{C}_{\ell} \rangle) ( \hat{C}_{\ell^{'}} - \langle \hat{C}_{\ell^{'}} \rangle) \rangle_{MC} \, .
\end{equation}
This approach has been widely used (see e.g. \citet{Jones:2006, QUAD2,Planck2013dpc}), and it is also the baseline procedure for {\it cROMAster} used for the analysis reported in this paper\footnote{For the sake of completeness, we notice that {\it cROMAster} implements also an approach for error estimation based
on bootstrapping: for a given $\ell$, we generate a set of fake $\tilde{a}_{\ell m}$ by resampling the observed ${a}_{\ell m}$ with uniform probability, and these are used to compute a set of fake $\tilde{C}_{\ell}$ to be fed into equation (\ref{eq:mccov}) to estimate the diagonal elements of the matrix. In order account for the sky cut, only $(2\ell+1)f_{eff}^{sky}$ elements of the $\tilde{a}_{\ell m}$ are averaged to compute the fake $\tilde{C}_{\ell}$. This is a very fast procedure based only on real data, and it is especially useful for testing purposes (i.e. when having a quick look to the data or checking for instrumental systematic effects). However, it is not accurate enough for cosmological analysis especially at low multipoles and when only a small fraction of the sky is considered. For this reason, we do not
employ it here.}.

Estimating off-diagonal elements of the covariance matrix with high accuracy requires a huge number of simulations. In fact, assuming that the covariance estimate is Wishart distributed, the uncertainty on
the diagonal elements simply scales as $\sqrt{(2/N_{sim})}$, while for the off-diagonal elements it scales as $\sqrt{1/N_{sim}}\times\sqrt{(Cov_{\ell \ell}*Cov_{\ell^{'} \ell^{'}})}/Cov_{\ell \ell^{'}}$, which is significantly worse 
in the presence of small correlations.

\citet{xspect} proposed an analytical approximation involving measured auto- and cross-power spectra which is accurate only at high multipoles and for large sky fraction. {\it Planck} power spectrum covariance is also based on an analytical approximation \citep{Planck:2013kta}, that apart from being rather more complex, involves a fiducial model $C_{\ell}$ as well as assuming uncorrelated pixel noise. 

\section{Computational requirements}
\label{computationalrequirements}

{\it BolPol} is a fully parallel implementation of the QML method, as described in Section \ref{bolpol}, written in Fortran90. Since the method works in pixel space the computational cost rapidly increases as one considers higher resolution maps of a given sky area. The original code, that has been applied to WMAP 5 year data in \citep{Gruppuso2009,Paci:2010wp}, to WMAP 7 year data in \citep{Gruppuso:2010nd,Paci:2013gs}, and to WMAP 9 year data in \citep{Gruppuso:2013xba}, performs the analysis both in temperature and polarization. Here we consider a modification of such implementation that works only on the temperature sector and can therefore ingest higher resolution maps than the original code. 
The following computational requirements are referred to this version of {\it BolPol} which has been already applied to {\it Planck} data \citep{Planck:2013kta,Ade:2013sta}.

The inversion of the covariance matrix {\bf C} scales as the third power of the side of the matrix, i.e. $\mathpzc{O}(N^3)$ being $N$ the number of observed pixels. The number of operations is roughly driven, once the inversion of the total covariance matrix is done, by the matrix-matrix multiplications to build the operators ${\bf E}_{\ell}^{X}$ in equation (\ref{El}) and by calculating the Fisher matrix $F_{XX'}^{\ell \ell'}$ given in equation (\ref{FF}). 
The memory, RAM, required to build these matrices is of the order of $\mathpzc{O}(\Delta \ell N^2)$ where $\Delta \ell$ is the range in multipoles of ${\bf C}^{-1} (\partial {\bf C}/\partial C_{\ell}^X)$ (for every X) that are built and kept in memory during the execution time. This implementation is memory demanding, but it has the benefit of speeding up the computations.
For a temperature all-sky map of 49152 pixels (resolution parametrized by the HEALPix\footnote{http://healpix.jpl.nasa.gov/} parameter $N_{side}=64$\footnote{For the reader who is not familiar with HEALPix library, we remind that $N_{side}$ is implicitly defined by  $N_{pix}=12 \, N_{side}^2$, where $N_{pix}$ is the total number of pixels in a CMB map.}, see \cite{gorski}), the total amount of RAM needed by {\it BolPol} 
is 19 TB. When we run the code into massively parallel computer clusters such as FERMI, at CINECA\footnote{http://www.cineca.it/}, it takes roughly one day using 16384 cores to perform a MC of 1000 maps with 20\% of the sky masked.
Possible future development and optimization of the code will lead the QML method to analyse higher resolution maps, but, at the moment, the QML method can be applied only to low resolution, i.e. large angular scales. 

On the other side {\it cROMAster} is very light and can be easily run on a common laptop since it works in harmonic space.

\section{Comparison}
\label{comparison}

The main idea of the paper is to compare the two aforementioned methods for APS extraction at the largest angular scales under realistic conditions. Our goal is to quantitatively find out to what extent it is worth to use a QML method with respect to a quicker and lighter pseudo-$C_{\ell}$ approach.

\subsection{Details of the simulations}

We consider several cases summarized in Table 1. The low resolution case, namely case 1 is parametrized by $N_{side}=64$ and corresponds to the maximum resolution that {\it BolPol} is currently capable to treat.
The high resolution case, i.e. case 2, with $N_{side}=256$ is analysed by {\it cROMAster}. 
Of course {\it cROMAster} can be run also at higher resolution but this does not impact the range of multipoles where we wish to compare the two codes which essentially is $\ell < \ell_{dec} \sim 90$.
The noise value of $1 \, \mu$K$^2$ at $N_{side}=256$ is the typical variance for a {\it Planck}-like experiment \citep{Ade:2013xsa}. 
The adopted value at large scale is still $1 \, \mu$K$^2$, that of course corresponds to an higher noise. 
This is done to regularize the numerics for this case. In practice, even if different, the noise levels for both cases of Table 1 are negligible and this different treatment 
does not impact on the analysis we perform.

\begin{figure}
\centering
\includegraphics[scale=0.25]{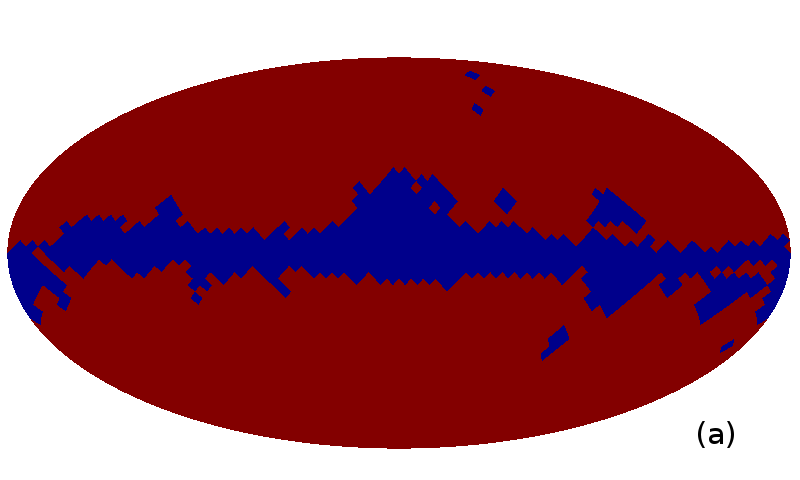}
\includegraphics[scale=0.33]{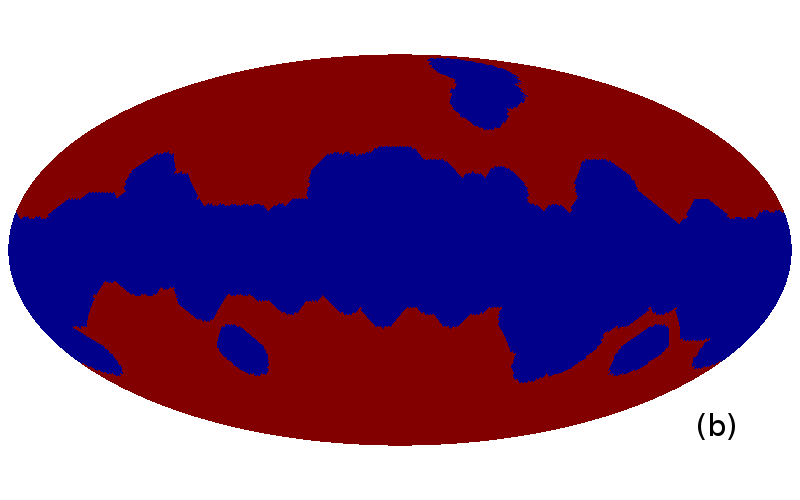}
\caption{ The masks used for the analyses in this paper. (a) WMAP kq85 temperature mask. (b) WMAP kq85 mask extended by 8 degrees.}\label{masks}
\end{figure}

Each of the cases of Table 1 is analysed with and without Galactic masks.
For the latter option, we used the WMAP kq85 temperature mask (Fig. \ref{masks}a), publicly available at the LAMBDA website\footnote{http://lambda.gsfc.nasa.gov/}, and an extended version of the kq85 that we call ``kq85 + 8 deg''(Fig. \ref{masks}b) where we have extended the edges of the mask by 8 degrees. The former mask excludes about $20\%$ of the sky, the latter about $48\%$.
In the full sky case and in a signal dominated regime, the two codes are algebraically equivalent.
We have checked that this is indeed the case for internal consistency\footnote{Since the codes run at different resolutions, their difference is driven by the considered smoothing, i.e. FWHM.}. 
Since a Galactic masking is in practice unavoidable in CMB data analysis we do not report on this unrealistic case.

For case 1 and 2 we analyse 1000 ``signal plus noise'' MC simulations where the signal is randomly extracted through the {\tt synfast} routine of the HEALPix package \citep{gorski}, from a $\Lambda$CDM {\it Planck} best fit model \citep{Planck:2013kta}
and the noise from a Gaussian distribution with variance given by the value reported in Table 1.

\begin{table}
\centering
\label{MCdetails}
\begin{tabular}{ccccccc}
\hline
Case & Res & Beam & Noise & $N_{sims}$ & Code & Masks\\
\hline
& $N_{side}$ & deg & $\mu$K$^2$ & \phantom{a} & & \\
\hline
1 & 64  & 0.916 & 1.0 & 1000 & B,C & a,b\\
2 & 256 & 0.573 & 1.0 & 1000 & C & a,b\\
\hline
\end{tabular}
\caption{Details of the considered MC simulations. 
Two cases are taken into account, each of them consist of 1000 realizations from the best fit of {\it Planck} model (in fact the choice of the model is irrelevant for our purposes).
First column ``Case'' is for type of simulations. Second column ``Res'' is for the resolution which is expressed in terms of the parameter $N_{side}$. Third column ``Beam'' is for the adopted FWHM. Fourth column is for the level of ``White noise'' which is given in $rms^2$. Fifth column ``$N_{sims}$'' contains the number of considered simulations. Sixth column ``Code'' specifies which code is applied for each case, where B stands for {\it BolPol} and C for {\it cROMAster}. Seventh column ``Masks'' identify the masks that have been applied to the simulated maps to perform the analysis: the WMAP kq85 maks (a) and the kq85 enlarged by 8 degrees (b). See the text for more details.}
\end{table}

\begin{figure}
\centering
\includegraphics[width=8cm]{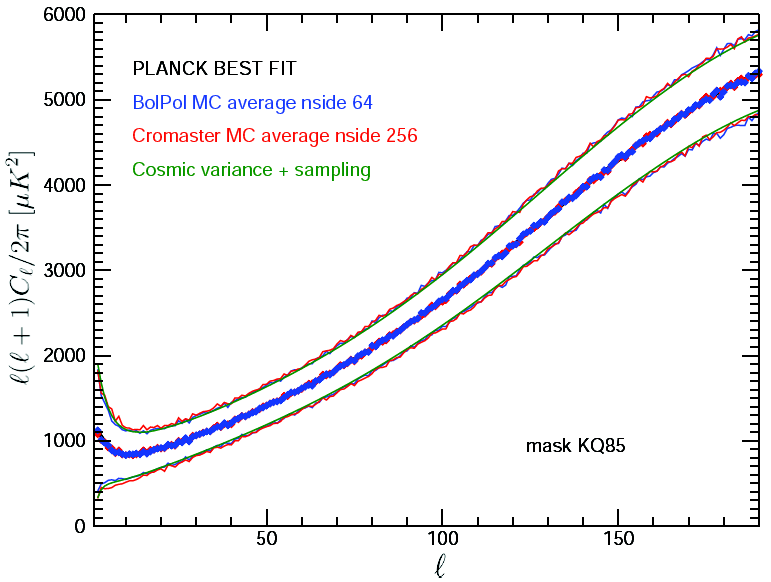}
\includegraphics[width=8cm]{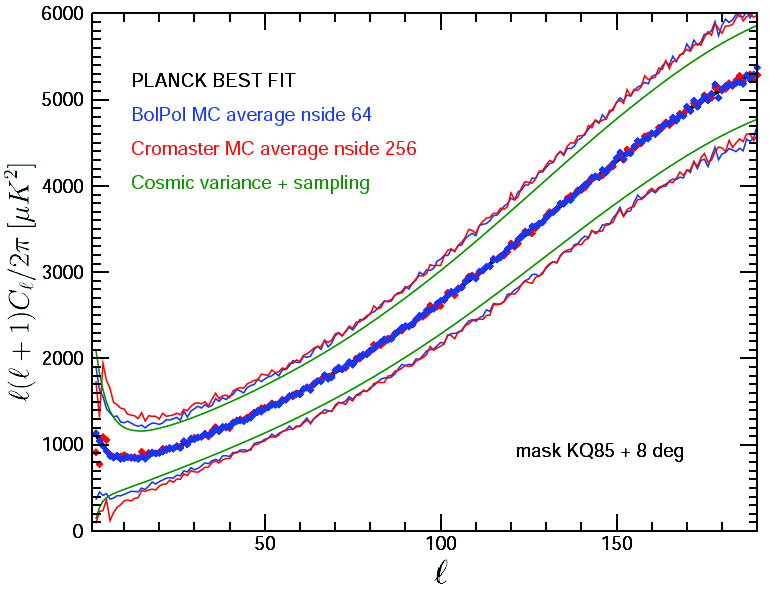}
\caption{Average and variance of the APS of the MC simulations analyzed  with {\it BolPol} (blue estimates) and {\it cROMAster} (red estimates) obtained masking the sky with the WMAP kq85 mask ({\it upper panel}) and the kq85 enlarged by 8 degrees ({\it lower panel}). See Table 1 for details.}
\label{MC results mask}
\end{figure}

The averages and variances of the APS of the two MC simulations are plotted in Fig. \ref{MC results mask} where we considered the WMAP kq85 mask ({\it upper panel}) and the kq85 mask enlarged by 8 degrees ({\it lower panel}).
These figures are considered as the validation of the performed extractions\footnote{In fact Fig. \ref{MC results mask} might be seen as the validation of both codes and extractions at the same time. 
However, an extensive validation of the codes is already given in \citep{Gruppuso2009} and in \citep{polenta}.}. 

\subsection{Figure of Merit}

In order to make a detailed comparison between the two methods, we have to define a suitable estimator.
Our approach is very similar to what is proposed by \cite{efstathiou}.

\begin{figure}
\centering
\includegraphics[scale=0.38]{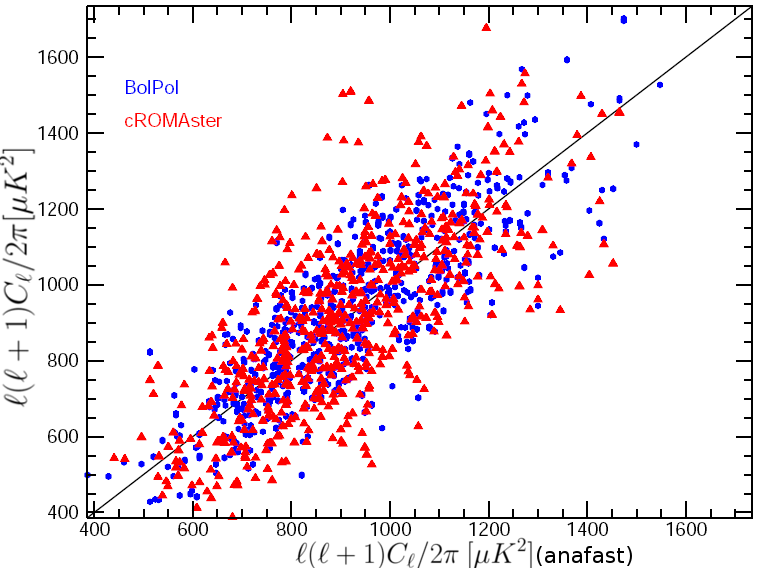}
\includegraphics[scale=0.38]{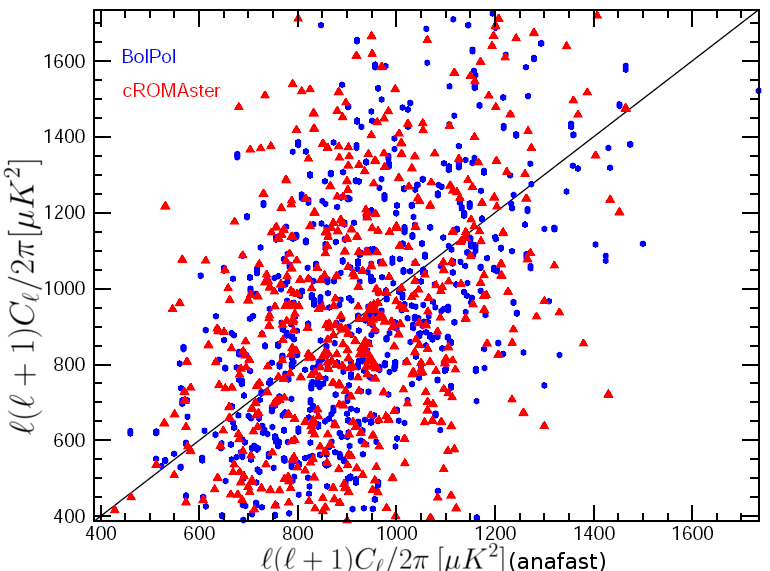}
\caption{All the $1000$ APS estimates at multipole $\mathbf{\ell=20}$ for the masked case of Table 1 when we considered the WMAP kq85 mask ({\it upper panel}) and the kq85 mask enlarged by 8 degrees ({\it lower panel}). Blue circles estimates are for {\it BolPol} and red triangles estimates for {\it cROMAster}.}
\label{confronto anafast}
\end{figure}

For each multipole and for each realization of MC simulations, we compute the APS and build plots as in Fig. \ref{confronto anafast}. 
In such a figure each point $P=(x,y)$ has the abscissa $x$ given by the APS obtained with {\tt anafast}\footnote{It is an HEALPix routine \citep{gorski}.} in the ideal case\footnote{In such an ideal case {\tt anafast} provides the true APS of the maps.} (i.e. full sky and no noise) and the ordinate $y$ that is given by the APS estimated through the {\it BolPol} or {\it cROMAster} for the cases of Table 1.

If the codes were ``perfect'' only the diagonal of this kind of plots would be populated (see black solid line in Fig.~\ref{confronto anafast}).
In fact there are two clouds of points, one for {\it BolPol} estimates, shown in blue, and one for the {\it cROMAster} estimates, shown in red.
The idea is to measure the dispersion of the two clouds around the solid black line. 
This defines our estimator aimed at the comparison of the two codes. 
The code that shows larger dispersion has an intrinsic larger variance in the determination of the APS.
In practice, for each single multipole $\ell$ we define the variance $D_{\ell}^2$ as the mean of the squared distance $d_{\ell}$ of each point $P$ from the line $y=x$, which is the diagonal of the first quadrant of this Cartesian plane,

\begin{equation}
D_{\ell}^{2 [B/C]} \equiv \langle d_{\ell}(P^{[B/C]},diagonal)^2 \rangle
\, ,
\label{figureofmeritdef}
\end{equation}
where the labels $^{B/C}$ refer to {\it BolPol} and {\it cROMAster} and with $\langle ... \rangle$ standing for the ``ensamble'' average. We underline that, in this way, the estimator cancels the uncertainty due to the cosmic variance that is the same for both the codes and highlight their different intrinsic variance.
Taking the square root of equation (\ref{figureofmeritdef}) we obtain

\begin{equation}
D_{\ell}^{[B/C]} = {\ell (\ell +1) \over {2 \, \sqrt{2} \, \pi}} (\langle (C_{\ell}^{[B/C]}-C_{\ell}^{A})^2 \rangle)^{1/2}
\, ,
\label{figureofmeritdef2}
\end{equation}
where $C_{\ell}^{A}$ is the APS computed with {\tt anafast} in the ideal case.
From equation (\ref{figureofmeritdef2}) it is clear that the unit of $D_{\ell}^{B/C}$ is the same as the one used for the APS, that in our case is $\mu$K$^2$. 
Equation (\ref{figureofmeritdef2}) is what we consider in the next section to perform the comparison.

\begin{figure*}
\centering
\includegraphics[width=8.4cm]{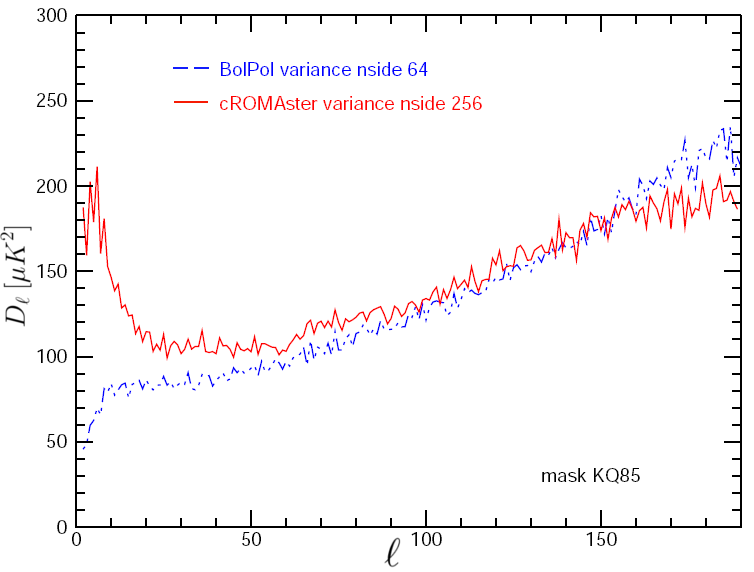}
\includegraphics[width=8.4cm]{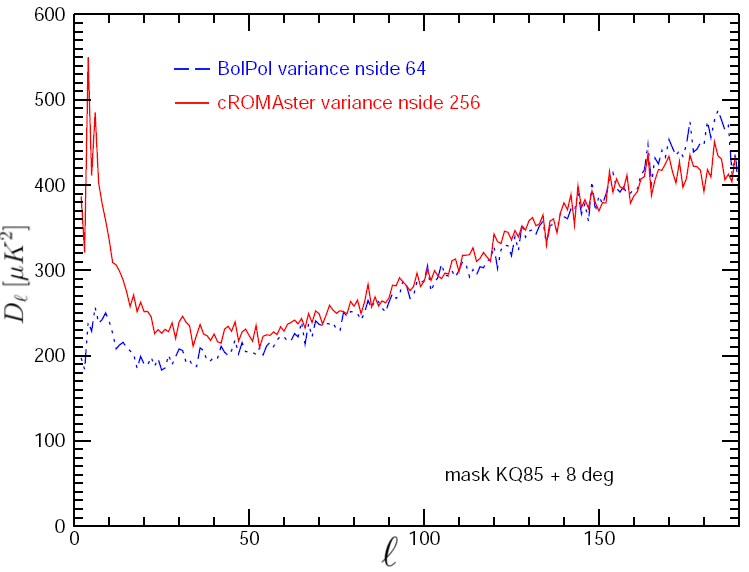}
\caption{$D_{\ell}$ vs multipole $\ell$ for each of the cases described in Table 1, when we considered the WMAP kq85 mask ({\it left}) and the kq85 mask enlarged by 8 degrees ({\it right}). Solid line for {\it Cromaster}, dashed line for {\it BolPol}.} 
\label{variancemaskedsky}
\end{figure*}

\subsection{Results}
\label{results}

\begin{figure}
\centering
\includegraphics[width=8.5cm]{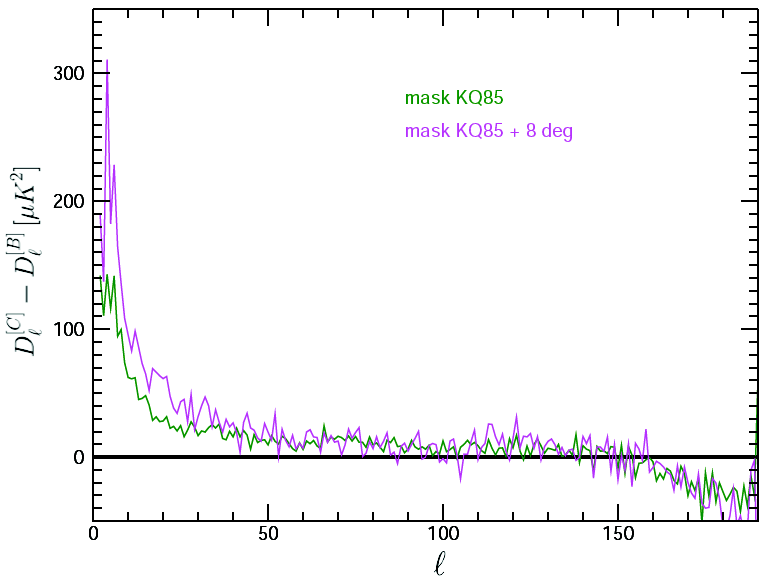}
\caption{$\mathbf{D_{\ell}^{[C]}-D_{\ell}^{[B]}}$ vs $\ell$ when we considered the WMAP kq85 mask ({\it green}) and the kq85 mask enlarged by 8 degrees ({\it magenta}).}
\label{varianzarelativamaskedsky}
\end{figure}

Fig. \ref{variancemaskedsky} shows the estimator $D_{\ell}$, defined in equation (\ref{figureofmeritdef2}), as a function of the multipole $\ell$ for each of the cases of Table 1.
This plot demonstrates that the intrinsic variance of {\it BolPol} is lower than the intrinsic variance of {\it cROMAster} up to $\ell \sim 100$.
The differences between the two estimators, $(D_{\ell}^{[C]}-D_{\ell}^{[B]})$ versus $\ell$ is shown in Fig. \ref{varianzarelativamaskedsky}.
This makes clear that the difference in the accuracy of the two methods is higher at lowest multipoles and that it grows as the number of masked pixel increases. 
In particular, when we consider the WMAP kq85 mask (kq85 enlarged by 8 degrees), the intrinsic dispersion introduced by the pseudo-$C_{\ell}$ method is {\it at least} a factor of $3$ (a factor of $2$) greater than that of the QML estimates for $\ell \leq 10$. In the range $20 \leq \ell \leq 60$ the QML is about $20\%$ ($15\%$) more accurate than the pseudo-$C_{\ell}$ method.

\begin{figure}
\centering
\includegraphics[width=8.5cm]{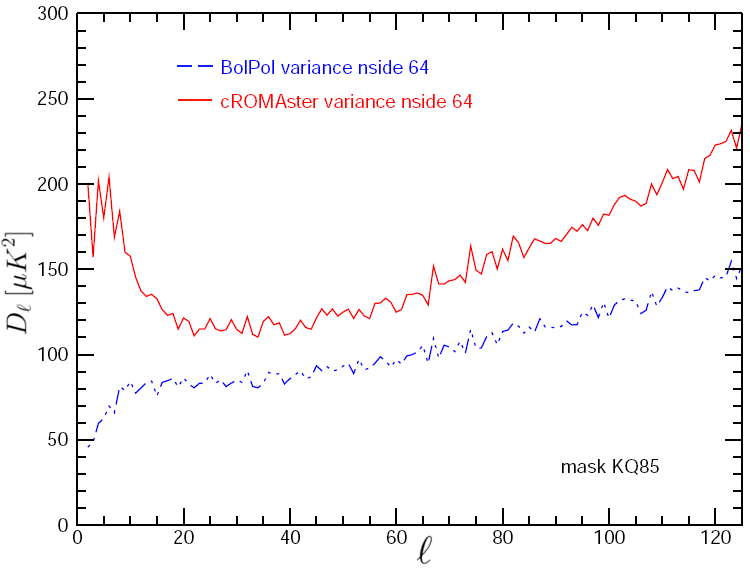}
\caption{$D_{\ell}$ vs multipole $\ell$ as in Fig. \ref{variancemaskedsky}({\it left}), but with the codes run at the same resolution ($N_{side}=64$).}
\label{variancemaskedskylowres}
\end{figure}

At higher multipoles, i.e. $\ell > 100$, the larger QML intrinsic variance displayed in Fig. \ref{variancemaskedsky} is entirely due to the lower resolution at which {\it BolPol} is run with respect to {\it cROMAster}. Note however that when the two codes are run at the same resolution, i.e. $N_{side}=64$, the QML has always a smaller variance than {\it cROMAster} in the commonly valid multipole domain\footnote{{\it cROMAster} results are fully reliable only up to $\ell_{max}=2 \times N_{side}=128$}, as shown in Fig. \ref{variancemaskedskylowres} when we used the WMAP kq85 mask\footnote{We obtained the same result when we considered the enlarged mask.}. These results show that the resolution given by $N_{side} = 64$ is enough to have an optimal APS extraction with the QML method in the range of interest ($\ell < \ell_{dec} \sim 90$) compared to pseudo-$C_{\ell}$ estimates performed on maps at the best resolution allowed by the observations.

\subsection{Applications}
\label{WMAP}

We consider here two estimators of anomalies used at large scales in CMB data analysis to illustrate the benefit of applying an optimal APS estimator. The estimator, $R$, see \cite{Kim e Naselsky 2010} and \cite{Gruppuso:2010nd} for the TT parity analysis, is $R\equiv C_{+}^{TT}/C_{-}^{TT}$ where $C_{+/-}^{TT} \equiv 1/(\ell_{tot}(+/-)) \sum_{\ell=2,l_{max}}^{+/-} \ell(\ell+1)/(2\pi) C_{\ell}^{TT}$ and $\ell_{tot}(+/-)$ is the total number of even (+) or odd (-) multipoles taken into account in the sum. The Variance estimator, $\sigma^2$, (e.g. \cite{Monteserin2008}, \cite{Cruz2010}, \cite{Ade:2013sta}, \cite{Gruppuso:2013xba} and reference therein) is defined by $\sigma^2 = \langle \delta T^2 \rangle=\sum_{\ell \geqslant 2}^{\ell_{max}} \left(2\ell+1/(4\pi)\right) C_{\ell}^{TT}$.

\begin{figure}
\centering
\includegraphics[width=8.5cm]{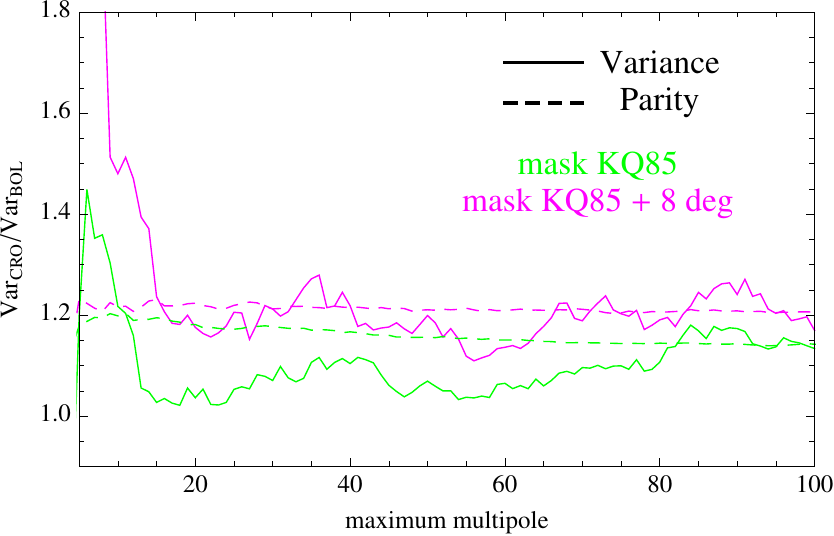}
\caption{Ratio of the variances ({\it cROMAster} over {\it BolPol}) for the Variance (solid line) and the TT Parity (dashed line) estimators as a function of the multipoles obtained from the {\it BolPol} and {\it cROMAster} APS extracted in the previous section when we considered the WMAP kq85 mask ({\it green}) and the kq85 mask enlarged by 8 degrees ({\it magenta}).}
\label{parity&variance}
\end{figure}

The variances of these two estimators are affected by the uncertainties on the APS. In Fig.\ref{parity&variance} we show the ratio of the variances of $R$ obtained through the APS extracted by {\it cROMAster} and {\it BolPol} for each $\ell_{max}$. In the same figure, we show also the ratio between the variances of $\sigma^2$ obtained with {\it cROMAster} and {\it BolPol}.
It is clear that the lower uncertainty given by {\it BolPol} leads to a lower variance for both the two estimators in the range of interest ($\ell \leq \ell_{dec} \simeq 90$). 
For the TT Parity estimator the average gain in efficiency of about 10\% when we use the WMAP kq85 mask (20\% for the kq85 mask enlarged by 8 degrees) with a peak at the lowest scales becoming higher than 40\% for both the masks.
In the case of Variance estimator, when we consider the WMAP kq85 mask, the variance obtained from {\it BolPol} APS is lower than the one obtained from the {\it cROMAster} APS by a factor of about 15\% becoming even higher for very large scales ($\ell \leq 40$). When we consider the kq85 mask enlarged by 8 degrees the gain in accuracy, when we use the QML estimator, is always about 22\%.
We also note that since the Variance estimator strongly depends on very low multipoles where the APS extractor methods have their larger differences, its ratio is greater than 1 even at multipoles higher than $\ell_{dec}$.

\section{Conclusion}
\label{conclusions}

Our main result is given by  Fig. \ref{variancemaskedsky} and \ref{varianzarelativamaskedsky} where the intrinsic variance, see equation (\ref{figureofmeritdef2}), of the two APS estimators, namely {\it BolPol} and {\it cROMAster}, are compared under realistic conditions. We have found that the QML method is markedly preferable in the range $2 \leq \ell \leq 100$.
Moreover, we note that the largest difference between the two codes is for the lowest multipoles:
for $\ell$ smaller than  $\sim 20$ the square root of the intrinsic variance introduced in the estimates by the pseudo-$C_{\ell}$ is {\it at least} up to three times (two times) the QML one when we consider the WMAP kq85 mask (respectively the kq85 mask enlarged by 8 degrees).
For higher multipoles (i.e. $\ell \geq 100$) we observe an opposite behaviour. This stems from the smoothing of the input maps that in turn it is a consequence of the adopted resolutions.
Note however that when the two codes are run at the same resolution, i.e. $N_{side} = 64$, the QML has always a smaller variance than cROMAster in the commonly valid multipole domain, as shown in Fig. \ref{variancemaskedskylowres}.
%It is due only to the smoothing applied to the input maps that in turn it is consequence of different resolutions considered.

We have also analysed how the intrinsic variance of the two APS methods impacts on some typical large scales anomaly estimators like the TT Parity estimator and Variance estimator. In conclusion, the use of {\it BolPol} for low resolution map analysis will bring to tighter constraints for these kind of estimators.

Therefore we suggest to use the QML estimator and not the pseudo-$C_{\ell}$ method in order to perform accurate analyses that are based on the APS at large angular scales (at least $\ell \leq \ell_{dec} \simeq 90$).
This might be of particular interest for studying large scale anomalies in the temperature anisotropy pattern. In a future work we will extend this analysis to the polarization field, which is crucial to reveal the reionization imprints at large scale.

\section*{Acknowledgments}

We acknowledge the use of computing facilities at NERSC (USA) and CINECA (ITALY). 
We acknowledge use of the HEALPix (Gorski et al. 2005) software and analysis package for deriving the results in this paper. 
We acknowledge the use of the Legacy Archive for Microwave Background Data Analysis (LAMBDA), part of the High Energy Astrophysics Science Archive Center (HEASARC). HEASARC/LAMBDA is a service of the Astrophysics Science Division at the NASA Goddard Space Flight Center.
Work supported by ASI through ASI/INAF Agreement I/072/09/0 for the {\it Planck} LFI Activity of Phase E2
 and by MIUR through PRIN 2009 (grant n. 2009XZ54H2).

%\appendix

\label{lastpage}

\end{document}